\documentclass[aps,pra,twocolumn,reprint,superscriptaddress,amsmath,amssymb]{revtex4-1} % APS journal style

\usepackage{soul} % For highlighting
\usepackage{graphicx} % Include figure files
\usepackage{dcolumn}  % Align table columns on decimal point
\usepackage{bm}       % bold math
\usepackage[colorlinks,urlcolor=blue,citecolor=blue,linkcolor=blue,pdfstartview=FitH]{hyperref}
\usepackage{enumerate}
\usepackage{amsmath}
\usepackage{color}
\usepackage{cases}
\usepackage{microtype}
\usepackage{siunitx}
\usepackage{subdepth}
\usepackage{todonotes}
\usepackage[utf8]{inputenc}
\usepackage[T1]{fontenc}
\usepackage{txfonts} %PRA-like font (math)
\begin{document}

\title{Geometric phases of a vortex in a superfluid}

\author{Rodney E. S. Polkinghorne}
\affiliation{Optical Sciences Centre, Swinburne University of Technology, Melbourne 3122, Australia}
\author{Andrew J. Groszek}
\affiliation{ARC Centre of Excellence in Future Low-Energy Electronics Technologies, School of Mathematics and Physics, University of Queensland, St Lucia, QLD 4072, Australia}
\affiliation{ARC Centre of Excellence for Engineered Quantum Systems, School of Mathematics and Physics, University of Queensland, St Lucia, QLD 4072, Australia}
\author{Tapio P. Simula}
\affiliation{Optical Sciences Centre, Swinburne University of Technology, Melbourne 3122, Australia}

\begin{abstract}
We consider geometric phases of mobile quantum vortices in superfluid Bose--Einstein condensates. Haldane and Wu [Phys.~Rev.~Lett.~{\bf 55}, 2887 (1985)] showed that the geometric phase,  $\gamma_{\mathcal C}=2\pi N_{\mathcal C}$, of such a vortex is determined by the number of condensate atoms $N_{\mathcal C}$ enclosed by the vortex trajectory. Considering an experimentally realistic freely orbiting vortex leads to an apparent disagreement with this prediction. We resolve it using the superfluid electrodynamics picture, which allows us to identify two additional contributions to the measured geometric phase; (i) a topologically protected edge current of vortices at the condensate boundary, and (ii) a superfluid displacement current. Our results generalise to, and pave the way for experimental measurements of vortex geometric phases using scalar and spinor Bose--Einstein condensates, and superfluid Fermi gases.
\end{abstract}

\maketitle

The Aharonov--Bohm phase \cite{Aharonov1959a} and the Pancharatnam--Berry phase \cite{Pancharatnam1959a,berry_quantal_1984} are classic examples of an ever growing family of geometric phases that influence many areas of physics \cite{Cohen2019a,Wilczek1981a}. Topological invariants and geometric phases are closely related \cite{Nakaharabook,Thoulessbook} and in two-dimensional systems \cite{Leinaas1977} they have major consequences for prospecting future technologies such as dissipationless conductors beyond quantum Hall systems \cite{QHE} and topological quantum computers \cite{kitaev1997quantum,nayak2008non,pachos2012introduction,field2018a}.

Soon after Berry's seminal result \cite{berry_quantal_1984}, Haldane and Wu \cite{haldane_quantum_1985} considered a quantum vortex being adiabatically transported along a closed path $\mathcal C$ in a two-dimensional superfluid and conjectured that as a consequence the system acquires a geometric phase $\gamma_{\mathcal C}=2\pi N_{\mathcal C}$, where $N_{\mathcal C}$ is the number of atoms enclosed by the path of the vortex. Perhaps surprisingly, this profound result remains to be confirmed by experiments. Thouless, Ao, and Niu  further showed that the geometric phase of such a vortex is related to the transverse force on a vortex \cite{thouless_vortex_1993,thouless_transverse_1996}, which is further linked to the vortex mass in a superfluid \cite{thouless2007a,Zwierlein2014a,simula_vortex_2018,Simula2020a}.

Quantum vortices in superfluid helium are hard to control and interrogate due to their subnanometer-scale vortex core structure being beyond optical resolvability. By contrast, in atomic Bose--Einstein condensates (BECs) and superfluid Fermi gases, quantised vortices can readily be both guided and imaged using lasers. Indeed, vortices in BECs \cite{anderson_vortex_2000,Freilich2010a} and superfluid Fermi gases \cite{Zwierlein2014a} have been observed to undergo adiabatic periodic orbital motion \cite{virtanen_adiabaticity_2001} and are therefore well suited for designing experiments to study the geometric phase, transverse force, and the vortex mass in a superfluid.

Here we show that a straightforward application of the result by Haldane and Wu \cite{haldane_quantum_1985} to experimentally realistic superfluids leads to a seeming contradiction. To resolve this issue, we deploy a vortex electrodynamics description of two-dimensional superfluids and show that two additional sources of geometric phase emerge in such systems. These are (i) \emph{a topologically protected vortex edge current}, and (ii) \emph{a superfluid displacement current}. Properly accounting for these additional contributions, we recover the Haldane and Wu prediction, opening a pathway for experimental measurements of the vortex geometric phases in a superfluid.

Let us begin with a demonstration of the result derived by Haldane and Wu \cite{haldane_quantum_1985}. They considered a superfluid order parameter of the form $\psi({\bf r})=|\psi({\bf r})|e^{iS({\bf r})}$ with a real valued phase function $S({\bf r})=\arctan(x-x_v,y-y_v)$ that describes a quantised vortex in two-dimensional space whose position at time $t$ is parametrised by ${ R}(x_v,y_v,t)$. When the vortex phase singularity $R$ is transported along a closed path ${\mathcal C}$ that encircles $N_{\mathcal C}$ atoms of the superfluid, the geometric phase
\begin{equation}
\gamma_{\mathcal C}= i\oint_{\mathcal C}\langle n;{ R}|\nabla_{ R}|n;{ R}\rangle dR=2\pi N_{\mathcal C},
\label{Berryphase}
\end{equation}
where $|n;{ R}\rangle$ is an eigenstate parametrised by the vortex position $R$ \cite{haldane_quantum_1985}. Using Stokes' theorem the line integral in Eq.~(\ref{Berryphase}) may alternatively be expressed as an areal integral $\gamma_{\mathcal C}= \int_{\mathcal A} {\boldsymbol \Omega_n({ R})} \cdot d{\bf a}$, in terms of the Berry curvature
\begin{equation}
\Omega_n({ R})= \nabla_{R}\times {\bf A}_n({ R}),
\label{Berrycurvature}
\end{equation}
where ${\bf A}_n= i \langle n;R|\nabla_{R}|n;R\rangle$ is the Berry connection. We briefly mention the subtlety of gauge invariance of the Berry phase, which is not strictly satisfied in this vortex problem since in this formulation the vortex state $\psi$ is not an eigenstate of the superfluid evolution operator due to the embedded dynamics of the Bogoliubov phonons. Nevertheless, we proceed by assuming an approximate gauge invariance and associate the macroscopic condensate wavefunction $\psi({\bf r};R)$ with the eigenstate $|n;R\rangle$ in Eq.~(\ref{Berryphase}).

We first consider a method of imprinting the vortex around a closed trajectory, see the first column in Fig.~\ref{fig:F1}, the details of which can be found in the Supplemental Material \cite{supplement}. This yields a vortex geometric phase in agreement with the prediction by Haldane and Wu \cite{haldane_quantum_1985}, and this result is compared with a direct simulation of the Gross--Pitaevskii equation (GPE) in the second column. Frames \ref{fig:F1}(a) and (b) show the probability density $|\psi({\bf r};R)|^2$ mapped onto the color intensity with black corresponding to zero density and the maximum density occurring near the centre of each image. The location $R$ of the vortex core is visible as a black hole. The white circle shows the trajectory $\mathcal{C}$ of the vortex, initially placed at the location of the white marker. The colors in (a) and (b) correspond to different values of the phase function $S$ as marked on the edges of the image (a). Figure~\ref{fig:F1}(d) shows the
accumulated Berry curvature, Eq.~(\ref{Berrycurvature}), for the imprinted vortex, while the accumulated total geometric phase, Eq.~(\ref{Berryphase}), is shown in Fig.~\ref{fig:F1}(c) as a function of time (in units of the vortex orbital period) using green markers. The orange straight line,  $\gamma_{\mathcal C}(t)=t\omega_vN_{\mathcal C}$, is the prediction by Haldane and Wu \cite{haldane_quantum_1985}. The agreement between the green markers and the orange line is good and 
any small deviations
may be attributed to the broken axisymmetry of the condensate density around the vortex core \cite{supplement}. Varying the radius of the vortex trajectory, shown in Fig.~\ref{fig:F1}(f), further corroborates Eq.~(\ref{Berryphase}). The Supplemental Video SV1 \cite{supplement} shows that at the moment the vortex trajectory completes a full orbit, the negative contributions (cyan) in Fig.~\ref{fig:F1}(d) perfectly cancel out due to the positive contributions (red) from the diametrically opposite side of the vortex trajectory, and only the positive contribution to the Berry phase due to the atoms inside the vortex path remains.

\begin{figure}[!t]
    \centering
    \includegraphics[width=\columnwidth]{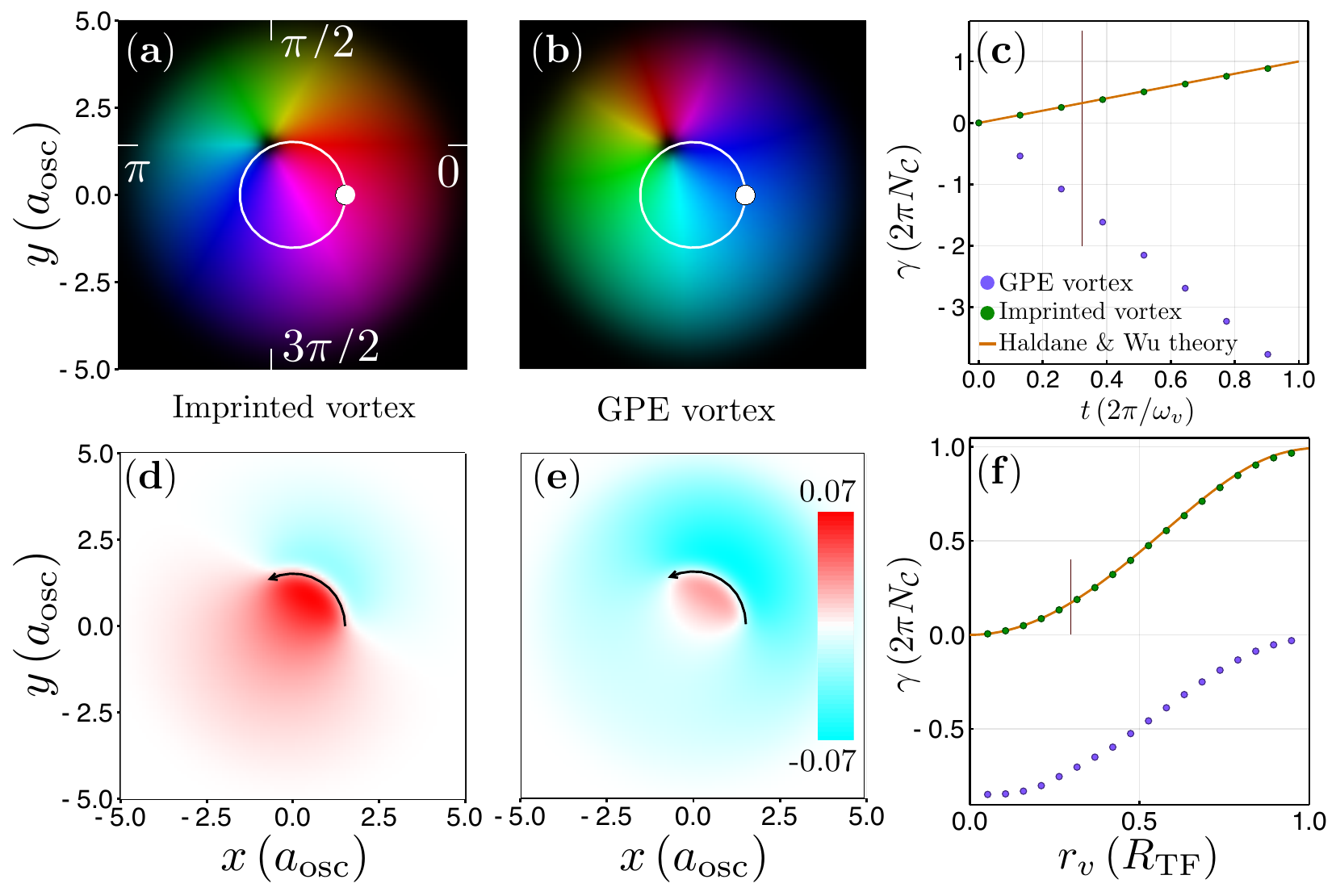}
    \caption{Geometric phase accumulated by a vortex orbiting in a harmonically trapped Bose--Einstein condensate. The condensate densities for an imprinted (a) and GPE evolved (b) vortex are identical and at $t=0$ the vortex is located at the position of the white markers. The subsequent vortex paths are show by the white circles. The colors correspond to the local value of the condensate phase and are quantified on the edges of frame (a). Frames (d) and (e) show the Berry curvature, Eq.~(\ref{Berrycurvature}), corresponding to (a) and (b), respectively. Frame (c) shows the accumulated geometric phase as functions of time, where $\omega_v$ is the angular frequency of the orbital motion of the vortex. The frame (f) shows the vortex geometric phase as functions of the vortex orbital radius $r_v$ in units of the Thomas--Fermi radius $R_{\rm TF}$ of the condensate, integrated over one complete vortex orbit. The vertical lines in (c) and (f) respectively mark the instant of time and the radial position of the vortices in (a,b,d,e). 
    Supplemental Videos \cite{supplement} SV1 and SV2  show the evolution of the snapshots in (a,b,d,e), and SV5 shows the hybridization of the kelvon and surfon quasiparticles when the vortex approaches the condensate surface \cite{simula_surfon_2002}.}
    \label{fig:F1}
\end{figure}

However, when the vortex position is allowed to freely evolve, as in the experiments \cite{anderson_vortex_2000,Freilich2010a,Zwierlein2014a}, a rather different result is obtained, shown by the purple markers in Figs.~\ref{fig:F1}(c) and (f). These results were obtained by numerically solving the time-dependent Gross--Pitaevskii equation 
\begin{equation}
    i\hbar\frac{\partial}{\partial t}\psi({\bf r}) =\left( {\hbar^2\over 2m}\nabla^2+V({\bf r})+g|\psi({\bf r})|^2 -\mu_0\right)\psi({\bf r}),
    \label{eq:GPE}
\end{equation}
for the case of a harmonic trap $V({\bf r})=m\omega_{\rm osc}^2 r^2/2$,  where $m$ is the mass of an atom, $g$ is the coupling constant, $\omega_{\rm osc}$ is the harmonic oscillator frequency that specifies the length scale $a_{\rm osc}=\sqrt{\hbar/m\omega_{\rm osc}}$. The total number of atoms in the condensate $N=\int |\psi({\bf r})|^2d^2{\bf r}$, and $\mu_0$ is the chemical potential of the vortex-free ground state. The details of the parameters and numerical implementations used for obtaining these results are provided in \cite{supplement}. Briefly, a vortex phase singularity is first imprinted at the location of the white marker in (b) and the state is relaxed in a rotating frame of reference to allow the vortex core structure to adjust to a self-consistent shape. The GPE is then propagated under the energy and angular momentum conserving GPE Hamiltonian in the laboratory frame, and as a result the vortex orbits around the trap centre along the shown trajectory at an approximately constant orbital angular frequency $\omega_v$, as observed in experiments \cite{anderson_vortex_2000,Freilich2010a,Zwierlein2014a}.

\begin{figure}[t]
    \centering
    \includegraphics[width=\columnwidth]{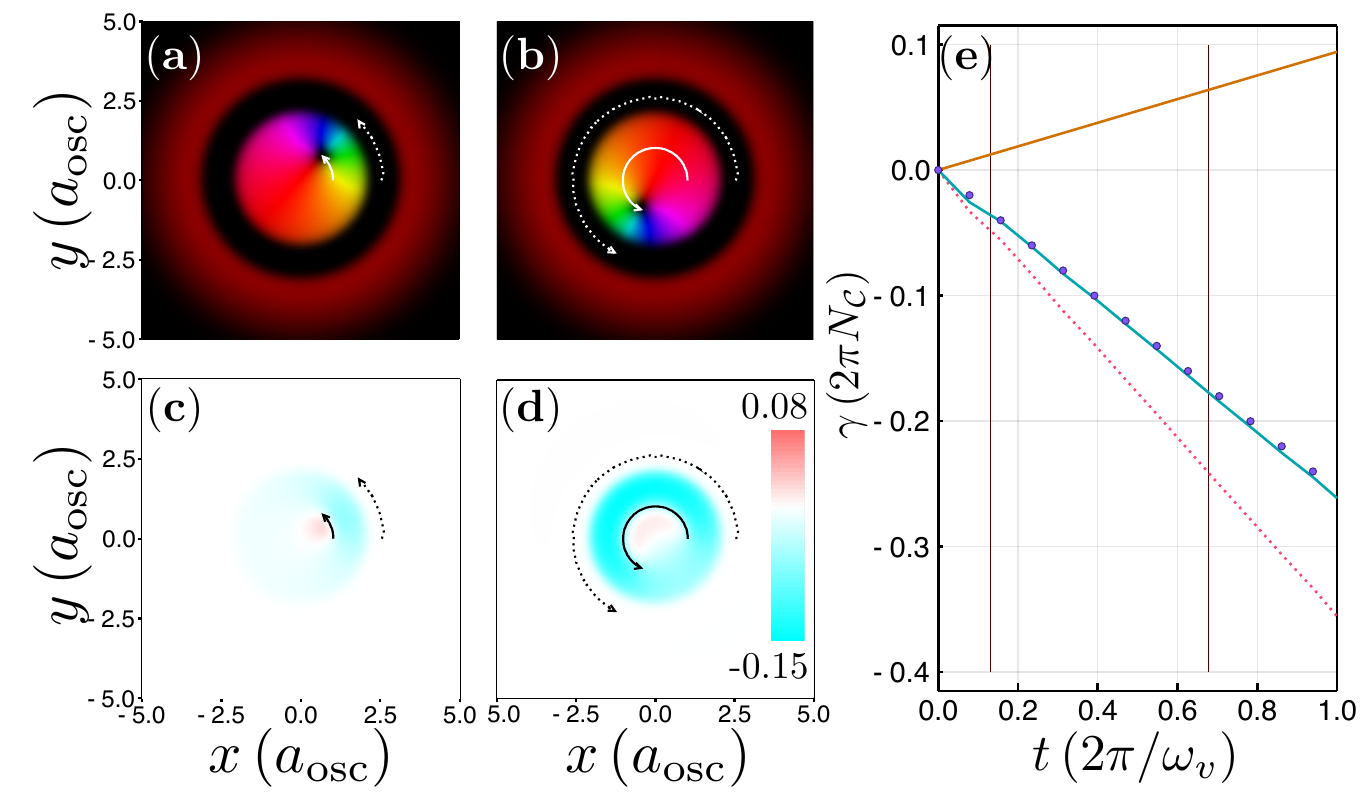}
    \caption{Topologically protected vortex edge current. (a) a Bose--Einstein condensate in a moat potential, Eq.~(\ref{moat}), with the paths of the visible inner vortex and the outer edge (anti)vortex inside the moat shown with white traces. (b) shows the system at a later time and, respectively, (c) and (d) show the accumulated Berry curvature, see also SV3 \cite{supplement}. Blue markers in (e) show the accumulated vortex Berry phase. The orange line with positive slope is the theory prediction \cite{haldane_quantum_1985} only accounting for the inner vortex, and the pink dashed line with a negative slope only accounts for the Berry phase due to the antivortex in the moat. The blue line sums up these two contributions due to the vortex and the moat antivortex and closely follows the measured total vortex Berry phase. The vertical lines indicate the times at which the snapshots in (a) and (b) were taken.}
    \label{fig:F2}
\end{figure}

Figure~\ref{fig:F1}(b) shows the phase-colored condensate density from the numerical GPE experiment at an instant of time marked in (c) with a vertical line. The accumulated Berry curvature shown in (e) is drastically different in comparison to the corresponding imprinted vortex case shown in (d). While the Berry phase in the imprinted vortex case counts the number of atoms inside the closed loop vortex path, the GPE case seems to count the number of atoms outside the vortex path (with a negative sign).

In order to understand the discrepancy between the imprinted vortex result, which agrees with the Haldane and Wu prediction \cite{haldane_quantum_1985}, and the GPE evolved one, which does not, we introduce a \emph{moat} potential of the form
\begin{equation}
V_{\rm moat}({\bf r}) = \frac{1}{2}m\omega_{\rm osc}^2 r^2 + a\hbar\omega_{\rm osc}\exp \left[ \frac{-(r-R_{\rm moat})^2}{2\sigma^2} \right],
\label{moat}
\end{equation}
where $a=100$ is a dimensionless moat amplitude, $\sigma=0.2\;a_{\rm osc}$ is the waist of the moat and $R_{\rm moat}=2.6\;a_{\rm osc}$ is the moat radius. The motivation for the moat is to engineer a controllable phase reference on the outside of the inner harmonically trapped part of the condensate. Similar potential landscapes were employed in the experiments by Eckel et.al. \cite{Eckel2014a}. The resulting condensate density and phase from the GPE are shown in Figs~\ref{fig:F2}(a) and (b) for two instants in time, quantified by the vertical lines in frame (e). The moat potential for this calculation is chosen such that there is no phase accumulation in the outer condensate annulus, which therefore defines a stable phase reference. The accumulated Berry curvatures corresponding to (a) and (b) are shown in (c) and (d), respectively. Similarly to Fig.~\ref{fig:F1}, the path of the inner vortex is shown in (a)-(d). In this case, a second vortex with opposite sign of circulation is found to reside within the moat \cite{supplement} and the path that it traces is shown in (a)-(d) by the outer dashed arcs. The presence of this antivortex in the moat is a topological necessity and we identify its motion as a topological \emph{vortex edge current}. We emphasise that the inner part of the BEC (inside the moat) is qualitatively equivalent (although having quantitatively different number of atoms) to the whole condensate simulated in Fig.~\ref{fig:F1}, where topological charge conservation also requires the existence of an antivortex (not shown) in the periphery of the system. 
 
 When the contributions to the Berry phase, shown in Fig.~\ref{fig:F2}(e), of both the inner vortex (orange line with a positive slope) and the antivortex in the moat (pink dashed line with a negative slope) are accounted for (blue line with a negative slope) an excellent agreement with the Haldane and Wu result \cite{haldane_quantum_1985} is recovered. The inner vortex covers a smaller area (small number $N_{\rm inner}$ of atoms) and accumulates a positive Berry phase. The moat vortex travels in the same direction as the inner vortex but has an opposite sign of circulation and therefore it accumulates a negative Berry phase whose magnitude is greater because the number of atoms $N_{\rm outer}$ it encircles is greater. The sum of these two contributions, due to the vortex and the (anti)vortex edge current, equals $-2\pi$ times the number of atoms, $N_{\rm annulus}=N_{\rm outer}-N_{\rm inner}$, within the (cyan) annulus in Fig.~\ref{fig:F2}(d) swept by the vortices. Note that in general, the inner vortex and the moat vortex do not travel at the same orbital angular frequency. 
 
This observation prompts an interpretation in terms of two-dimensional vortex electrodynamics \cite{Simula2020a}. In this picture the electric and magnetic fields are defined as
\begin{equation}
{\bf E}_{\rm sf}=\rho_s{\boldsymbol v}_{\rm s}\times {\bf e}_z
\hspace{5mm}
{\rm and}
\hspace{5mm}
{\bf B}_{\rm sf}= \frac{\hbar m}{g}\frac{\partial S}{\partial t} {\bf e}_z,
\label{EB}
\end{equation}
where $\rho_s=m|\psi({\bf r})|^2$, $\boldsymbol v_s=\hbar \nabla S /m$ is the superfluid velocity.
The superfluid Ampere--Maxwell law is
\begin{equation}
\nabla_\perp\times {\bf B}_{\rm sf}=\mu_v{\bf j}_v + \mu_v\epsilon_v \frac{\partial {\bf E}_{\rm sf}}{\partial t},
\label{eq:Ampere}
\end{equation}
where $\mu_v$ and $\epsilon_v$ are the superfluid vacuum constants and ${\bf j}_v$ is the vortex current density \cite{Simula2020a}. The full set of superfluid Maxwell equations are coupled with the exact vortex equation of motion \cite{groszek_motion_2018} that in the superfluid electrodynamics picture replaces the Lorentz force law. We then associate the Berry curvature with the superfluid magnetic field via
\begin{equation}
    { B}_{\rm sf}\; \hat=\frac{m}{2\pi}\; \Omega_n,
    \end{equation}
such that the closed loop Berry phase can be viewed as a superfluid magnetic flux 
\begin{equation}
  \gamma_\mathcal{C}=\frac{2\pi}{m}\int_{\mathcal A}  {\bf B}_{\rm sf}\cdot  d{\bf a} ,
    \end{equation}
where the mass of the atom is interpreted as the magnetic flux quantum.

The result in Fig.~\ref{fig:F2} was obtained using a judicious choice of the chemical potential $\mu_0$ such that the time derivative of the radial phase gradient across the moat, the last term $\mu_v\epsilon_v\partial_t {\bf E}_{\rm sf}$ in Eq.~(\ref{eq:Ampere}), is zero and the moat vortex orbital frequency equals that of the inner vortex. The Berry phase in this case accumulates solely due to the total vortex current ${\bf j}_v$ of the two vortices in the system. Extending this to arbitrary number of vortices yields 
\begin{equation}
\gamma_{\rm HW}=2\pi \sum_v {\rm sign}(v)N_v(v),
 \end{equation}
where $N_v(v)$ is the number of atoms swept across by the vortex $v$, and ${\rm sign}(v)$ is $+1$ for a vortex (antivortex) orbiting in the counter-clockwise (clockwise) direction and $-1$ for a vortex (antivortex) orbiting in the clockwise (counter-clockwise) direction. 

\begin{figure}[t]
    \centering
    \includegraphics[width=\columnwidth]{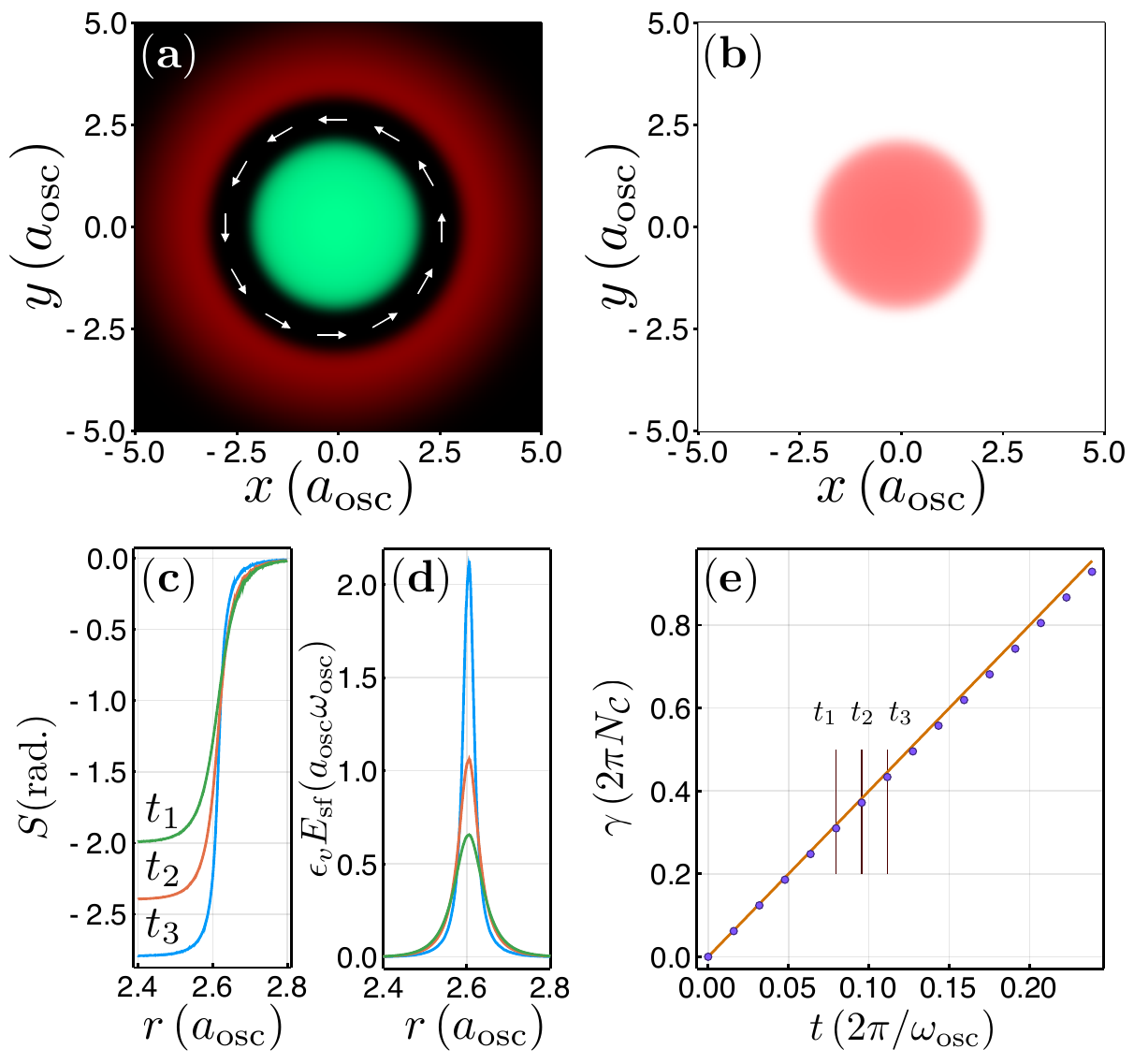}
    \caption{Superfluid displacement current. Frames (a) and (b) correspond to time $t\omega_{\rm osc}/2\pi=0.143$ and are similar to Fig.~\ref{fig:F2}(a) and (c), respectively, except that the vortices are absent here. The colorbar in Fig.~\ref{fig:F2}(d) applies to Fig.~\ref{fig:F3}(b) as well. The white arrows illustrate the superfluid displacement current that is localised in the moat. Frame (c) shows three samples of the evolution of the radial condensate phase across the (cylindrically symmetric) moat. Frame (d) shows the corresponding amplitudes of the superfluid electric field (the vertical axis has been scaled (divided) by a factor of 1000 for the sake of presentational clarity). The frame (e) shows the accumulated Berry phase as a function of time with the three vertical lines  specifying the sampling times used for the data  in frames (c) and (d). Supplemental Movie S3 shows the evolution of images (a)-(d) \cite{supplement}.}
    \label{fig:F3}
\end{figure}

To complete the correspondence with the superfluid Ampere--Maxwell law, Eq.~(\ref{eq:Ampere}), and to provide a geometric interpretation for $\mu_0$, we consider the case where the vortex current ${\bf j}_v$ is set to be zero. This corresponds to a vortex-free superfluid. Figure~\ref{fig:F3}(a) shows a ground state BEC in the moat potential where the effective local chemical potential in the condensate annulus outside the moat is zero. The inner part of the condensate has a nonzero chemical potential (with respect to $\mu_0$) such that there is a chemical potential difference in the radial direction across the moat. The result is a superfluid displacement current, the last term in Eq.~(\ref{eq:Ampere}), illustrated with white arrows in Fig~\ref{fig:F3}(a). Figures ~\ref{fig:F3}(b), (c) and (d) show, respectively, the accumulated Berry phase, the radial phase $S(r)$ at three different times, and the magnitude of the effective superfluid electric field, which is localised at the edge of the inner condensate, inside the moat. Figure~\ref{fig:F3}(e) shows the accumulated Berry phase as a function of time (blue markers) and the predicted superfluid displacement current (solid sloping line) according to the last term in Eq.~(\ref{eq:Ampere}). The three vertical lines correspond to the times sampled in (c) and (d). 

We note that as the condensate phase $S(t)$ inside the moat periodically cycles through its range $[-\pi,\pi)$, both the direction of the electric field and the rate of change of its amplitude change signs, resulting in a constant rate of Berry phase accumulation (effective magnetic flux) \cite{supplement}. In other words, the steadily changing condensate phase in the central region results in a uniform superfluid magnetic field $B_{\rm sf}$, which is understood to be induced by the superfluid displacement current, the last term in Eq.~(\ref{eq:Ampere}), localised within the moat as illustrated using the white arrows in Fig.~\ref{fig:F3}(a).

The agreement between the theory and the numerical experiments corroborate the vortex electrodynamics picture whereby the Berry curvature that is associated with the superfluid magnetic field has two sources: (i) a current of quantised vortices ${\bf j}_v$, and (ii) the displacement current $\partial {\bf E}_{\rm sf}/\partial t$ due to the spatio-temporal condensate phase evolution. Once both of these terms are accounted for, the apparent disagreement between the result of Haldane and Wu \cite{haldane_quantum_1985} for the vortex Berry phase and that predicted by time-evolution of the non-linear Schr\"odinger equation, see Fig.~\ref{fig:F1}, can be readily understood.

Our numerical results have verified that the prediction by Haldane and Wu, under the assumptions made in \cite{haldane_quantum_1985}, is correct. However, the way those assumptions apply to the boundary conditions of a realistic superfluid order parameter is quite subtle, and is obscured by the common but unphysical assumption of an infinite condensate of uniform density.  In particular, an infinitesimal perturbation to a ground state superfluid wave function with a chemical potential $\mu_0$ can transform a dynamical phase rotating uniformly at infinity as~$e^{ -i\mu_0 t / \hbar}$, into a geometric phase associated with an edge current of vortices with static condensate phase at infinity, or into any combination of dynamical and geometric phases depending on the specific details of the boundary condition that is realised. 

This ambiguity between the  geometric and dynamical phases in this interacting many-particle system bears similarity to the duality between particles and fields.  The traditional interpretation of a superfluid Bose--Einstein condensate is that the superfluid atoms are particles, and the vortices are a feature (akin to magnetic flux) embedded in the order parameter field, and that each atom encircling a vortex picks up a $2\pi$ phase winding.  But an alternative interpreation is possible, where the vortices are emergent particles, and the condensate wave function comprising the atoms is a dynamical gauge field that mediates the vortex interactions via phonons of the superfluid. Such particle-vortex duality has been extensively discussed within the context of condensed matter systems \cite{Peskin1978a,Dasgupta1981a,Lee1991a,Wang2015a,Metlitski2016a,Krach2016a,Mross2016a,Seiberg2016a}, where typically the particle substrate is electrons, and the geometric phase occurs due to Aharonov--Bohm type effects when charged particles loop around singular magnetic vector potentials.  In a superfluid, the dual picture assigns each vortex a charge equal to its quantised circulation, and the gradient and the rate of change of the order parameter phase are identified with effective electric and magnetic fields \cite{Simula2020a}.  When the results in Figs~\ref{fig:F2} and~\ref{fig:F3} are viewed through this window, the dynamics of the condensate phase is associated with a magnetic field whose flux is quantised with the mass of an atom corresponding to a quantum of flux, and the edge vortex is associated with a current of charged particles flowing around the superfluid magnetic field generated by the atoms that the vortices encircle. 

Interestingly, in the superfluid electrodynamics picture the vortex in a BEC, see Fig.~\ref{fig:F1}(a), realises a dynamic Corbino disk geometry \cite{Corbino1911a}, and opens a window for studies of a plethora of vortextronic applications familiar from electronic systems \cite{Onsager1952a,Kleinman1960a,Midgley1960a,Laughlin1981a,Bardyn2019a}. Considering two vortices initially located on diametrically opposite sides of the condensate, each of which then orbits half a circle, results in a \emph{vortex exchange phase} of $2\pi$ times an integer number of atoms, reflecting the bosonic origin of the vortices. These results also generalise to Fermi superfluids and spinor Bose--Einstein condensates. In the former case \cite{Zwierlein2014a,CoupledLattices2016a} Cooper-like pairing results in the effective vortex charge being $\kappa/2$ in contrast to the quantum of circulation $\kappa$ for the bosonic scalar superfluid, while in the latter case the fractional vortices are anyons that in general acquire a non-abelian exchange Berry phase upon braiding \cite{mawson2019a,field2018a}.  

In summary, we have studied geometric phases of a vortex in a superfluid having identified two new contributions: (i) a vortex edge current, and (ii) a superfluid displacement current. Our results open the path for experimental measurements of these geometric phases of vortices in superfluids.

\begin{acknowledgements}
This work was performed on the OzSTAR national facility at Swinburne University of Technology. The OzSTAR program receives funding in part from the Astronomy National Collaborative Research Infrastructure Strategy (NCRIS) allocation provided by the Australian Government.
The simulations made extensive use of the open source libraries Optim.jl~\cite{Mogensen2018} and DifferentialEquations.jl~\cite{rackauckas2017differentialequations}.
This research was funded by the Australian Government through the Australian Research Council (ARC) Discovery Project DP170104180 and the Future Fellowship FT180100020.
\end{acknowledgements}

\bibliography{ref_list}
\bibliographystyle{apsrev4-1}

\section{Supplementary material}

\section{Gross--Pitaevskii simulations}

The numerical simulations were implemented using a dimensionless form of the two-dimensional Gross--Pitaevskii equation (GPE),
\begin{equation}
    i\psi_t = {1\over2}\left(-\nabla^2 + r^2\right)\psi + C|\psi|^2\psi,
\end{equation}
where the length and time are expressed in units of the harmonic oscillator length $a_{\rm osc}=\sqrt{\hbar/m\omega_{\rm osc}}$, and the inverse trap frequency $\tau=1 / \omega_{\rm osc}$, respectively. The dimensionless condensate order parameter was normalised according to $\int |\psi|^2 dx\,dy=1$. In the two-dimensional formalism used here, the coupling constant is~$C=g N /(\hbar \omega_{\rm osc} a^2_{\rm osc} l_z)$ , where $l_z$~is the thickness of the sheet of condensate.  All simulations used~$C=500$.  For a $l_z=\SI{10}{\micro\meter}$~thick condensate made of~${}^{87}{\rm Rb}$, this would correspond to a condensate of approximately 77000~atoms in a radial harmonic oscillator trap of frequency $\omega_{\rm osc}=2\pi\times 10 $~Hz. The moat potential had an amplitude $a=100$, width $\sigma=0.2\;a_{\rm osc}$, and radius $R_{\rm moat}=2.6\;a_{\rm osc}$.

When initialising the orbiting vortices, to constrain a vortex at $r_v$, the order parameter was relaxed with a standard conjugate gradient routine.  After each step of relaxation, the value $\psi(r_v)$ was interpolated, and the order parameter was projected onto the subspace where $\psi(r_v)=0$.  The projection was performed by subtracting a Gaussian whose width was adjusted to cover the distance moved by the vortex in a relaxation step. When the trap rotation frequency matches the vortex orbital frequency, the relaxed vortex is stationary.  Therefore, as the relaxation converges, the subspace $\psi(r_v)=0$ becomes invariant under further relaxation, the projection ceases to have any effect, and the constrained order parameter is the same as the unconstrained one.

\section{Vortex imprinting method}
Vortex states $\psi_{\rm GPE}({\bf r},R)$ were first obtained by solving the GPE. For each location $R$ of the vortex along its orbit, an imprinted condensate wavefunction $\psi_{\rm imprinted}({\bf r},R)= |\psi_{\rm GPE}({\bf r},R)|e^{iS({\bf r};R)}$, where $S({\bf r};R)=\arctan(x-x_v,y-y_v)$ was constructed. In words, the GPE phase functions $S_{\rm GPE}$ were replaced by the cylindrically symmetric phase functions while retaining the true condensate densities. These vortex states were then used for calculating the geometric phases for the imprinted vortex case shown in Fig.~1 of the main text.

\begin{figure}[!t]
    \centering
    \includegraphics[width=\columnwidth]{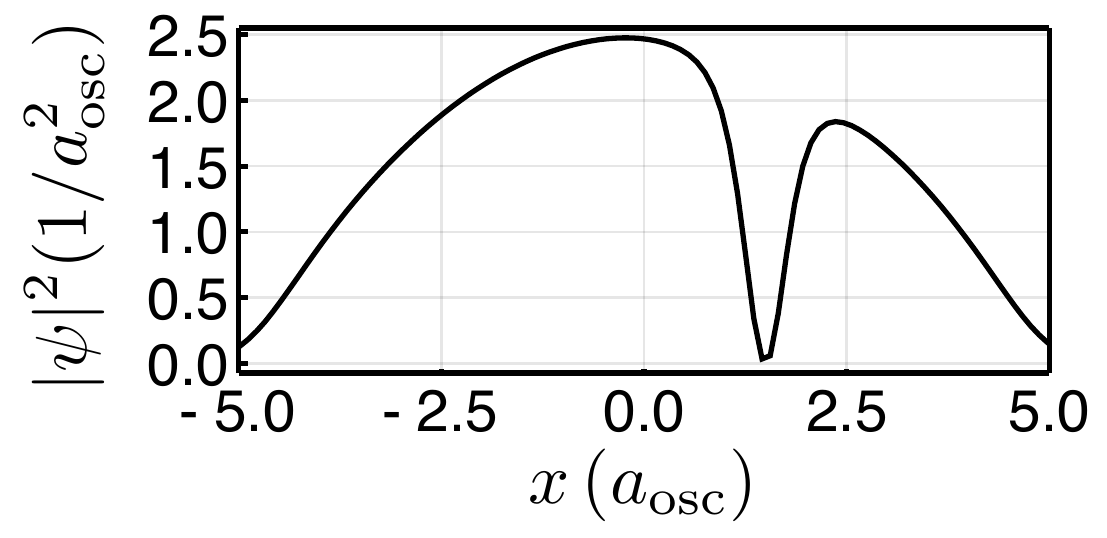}
    \caption{Condensate density along $x$-axis at $t=0$ in Fig.~1 of the main text. The vortex core is visible as the dip in the density.}
    \label{fig:FS1}
\end{figure}

\begin{figure}[!t]
    \centering
    \includegraphics[width=\columnwidth]{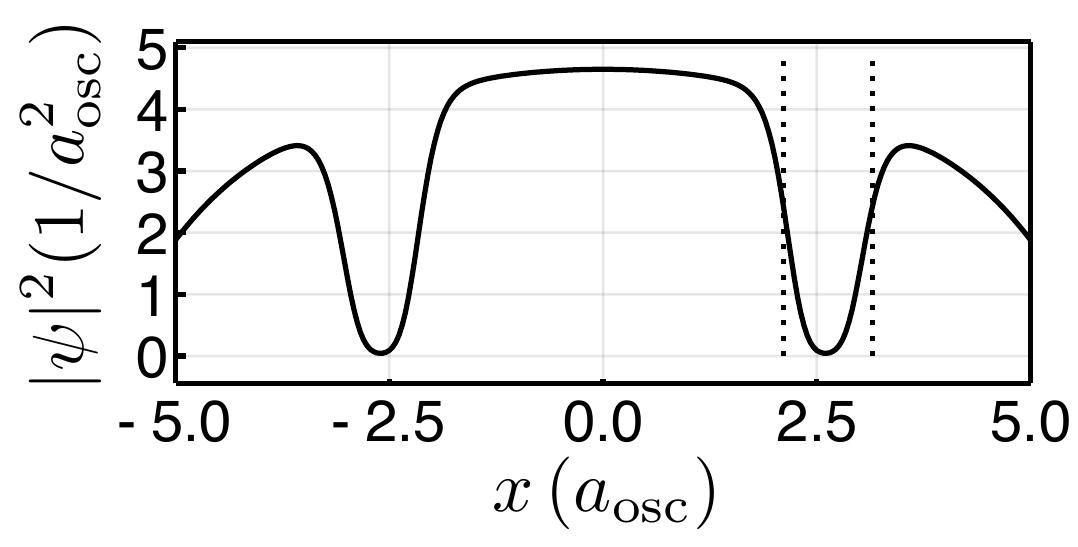}
    \caption{Condensate density along $x$-axis at $t=0$ in Fig.~3 of the main text. The moat is visible as the two dips in the density. The dotted lines show the radii at which the moat edges are plotted in Fig.~(\ref{fig:FS2}).}
    \label{fig:FS2}
\end{figure}

\section{Extracting the vortex Berry phase}

The vortex path~$\mathcal C$ is sampled in time as~$[R_1, R_2, \ldots]$, so the integral in Eq.~(1) of the main text can be approximated as
\begin{equation}
    \gamma_{\mathcal C} = \sum_j (\gamma_{j+1}-\gamma_j),
\end{equation}
where
\begin{equation}
    \gamma_{j+1}-\gamma_j = -{\rm Im}\langle n; R_{j+1}|n; R_j\rangle.
\end{equation}
For a condensate wavefunction~$\psi({\bf r}, R)$, this yields
\begin{equation}
\label{eq:bcurv}
    \gamma_j = \int \sum_{k=1}^{j-1}{\rm Im} \left(-\psi^\ast({\bf r},R_{k+1})\psi({\bf r},R_k)\right)\,d^2{\bf r}.
\end{equation}
The Berry phases shown in the main text are the~$\gamma_j$ obtained when the order parameter is substituted into Eq.~\eqref{eq:bcurv}.  All Berry phases are normalised by $N_{\mathcal C}=\int_{\mathcal A}|\psi({\bf r})|^2\,d^2{\bf r}$, where $\mathcal{A}$ is the area bounded by $\mathcal{C}$.  The Berry curvatures are the integrands
\begin{align}
\Omega_j=\sum_{k=1}^{j-1}{\rm Im} \left(-\psi^\ast({\bf r},R_{k+1})\psi({\bf r},R_k)\right).
\end{align}
Measuring the Berry phase requires knowledge of the condensate phase in addition to the condensate density that can be obtained by standard absorption imaging. This could be achieved for instance using propagation based phase retrieval methods such as the Gerchberg--Saxton algorithm or the Paganin's method. A proposed protocol would then be to (i) measure the interference patterns as in [26] (ii) extract the condensate phase using a phase retrieval method of choice, and (iii) use Eq.~\eqref{eq:bcurv}
to calculate the Berry curvature and phase.

\section{Vortex core asymmetry}

\begin{figure}[!t]
    \centering
    \includegraphics[width=\columnwidth]{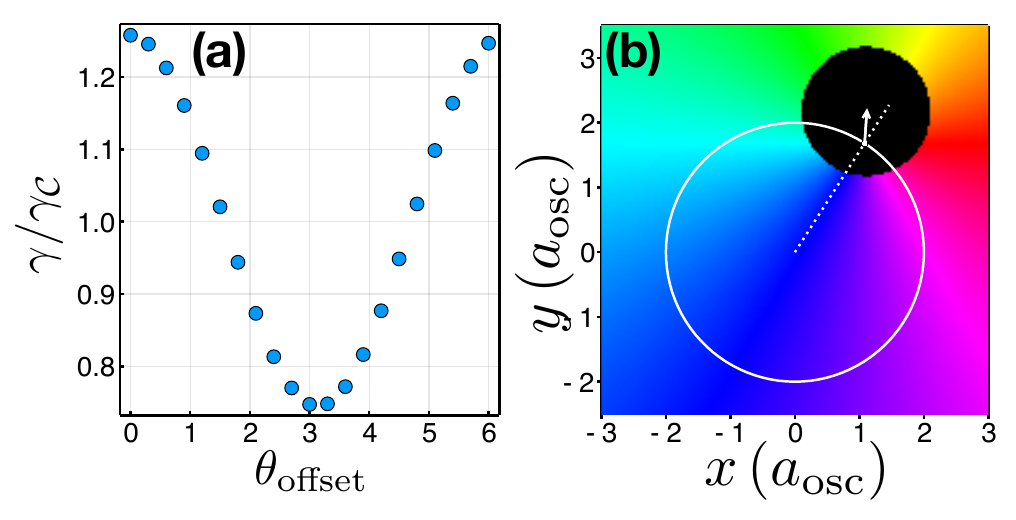}
    \caption{Effect of vortex core asymmetry on the geometric phase (a). Figure (b) shows a phase singularity, located at the intersection of the white circle and white dashed line. The order parameter has constant density, except in the region of the black disk of zero density covering the phase singularity. The core disk has radius $r_c=1.0\, a_{\rm osc}$, and its centre, located at the tip of the white arrow, is displaced by a distance $r_{\rm offset}=0.5\, a_{\rm osc}$ from the singularity.  The angle between the arrow and the dotted line, $\theta_{\rm offset}$, is the abscissa of Figure (a), which shows the geometric phase accumulated when the singularity orbits around the white circle, and the disk orbits rigidly on a concentric circle. A singularity with $r_c=0$ or $r_{\rm offset}=0$ accumulates Haldane \& Wu geometric phase of $\gamma = \gamma_\mathcal{C}$.}
    \label{fig:FS3}
\end{figure}

Figure~\ref{fig:FS3} shows the effect of an asymmetric density on the geometric phase accumulated by an orbiting vortex.  The vortex core has been displaced from the phase singularity. The effect is greatest when the core is displaced in the radial direction, along the white dashed line in Fig.~\ref{fig:FS3}(b), corresponding to the radial asymmetry that affects a vortex at the edge of the trap.

\section{Detecting the moat vortex}

Detecting the location of the inner vortex in Figs~1 and 2 of the main text is straightforward using standard numerical methods of measuring the phase winding. The presence and location of the vortex is also obvious from visual inspection of images (a) and (b) in Figs~1 and 2. Since the phase (dark red color) is constant in the annulus outside the moat, the topological charge conservation requires there to be an antivortex of opposite sign of circulation with respect to the inner visible vortex, somewhere within the moat region. However, determining its precise location is slightly more challenging in comparison of detecting the position of the inner vortex. The reason for this is that the whole moat may be viewed as a vortex core that has been stretched to an annular shape and therefore the phase singularity, in principle, could reside anywhere along the ring-shaped bottom of the Mexican hat-like moat potential. In reality, the constant phase contours converge to a singular point that defines the precise location of the moat vortex, shown in  Fig.~\ref{fig:FS4}. This point was determined numerically as follows:

\begin{figure}[!t]
    \centering
    \includegraphics[width=0.6\columnwidth]{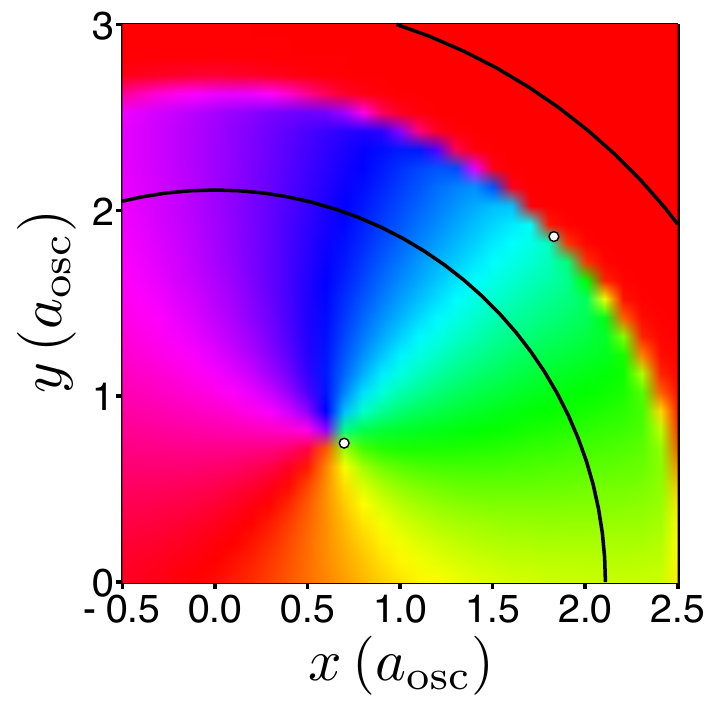}
    \caption{Vortex in the moat. The colors correspond to the local value of the condensate phase and the numerically measured locations of the inner vortex and the moat (anti)vortex are marked with small circular white markers. The black circles indicate the boundaries of the moat, where the condensate density is half of its maximum value in the outer annulus, as shown in Fig.~(\ref{fig:FS2}).}
    \label{fig:FS4}
\end{figure}

\begin{itemize}
    \item The usual way to detect unit-winding vortices has been to take the convolutions $P(w)=(z-w)^\ast\star\psi$ and $Q(w)=(z-w)\star\psi$, where $w=x+iy$ is a complex coordinate.  For a vortex, $|P(w)/\psi|$ has a sharp maximum at the singularity, and for an anti-vortex $|Q(w)/\psi|$ does.
    \item This fails as a vortex becomes more elongated.  In the limit of a soliton, both convolutions are zero.
    \item The first step is to approximately locate the moat vortex, which we do by finding the maximum of $|P+Q^\ast|/|\psi|$.  This relies on the order parameter being real outside the moat.  A form such as $|e^{i\phi}P+e^{-i\phi}Q^\ast|/|\psi|$ might be required in the general case.
    \item Every vortex or soliton with singularity at $r_v$ is a linear combination of $z-r_v$ and $(z-r_v)^\ast$, and therefore of 1, $z$ and $z^\ast$.  The moat vortex was located precisely by taking a least-squares fit $\psi=a+bz+cz^\ast$ at the pixels where $|P+Q^\ast|/|\psi|$ is largest, then solving for $r_v=(bc^\ast-a^\ast c)/(|a|^2-|b|^2)$.
    \item The least-squares method was also used to fit the inner vortices.  The convolutions are pixel-limited, while the spectral discretisations used in the simulations are continuous functions that can be interpolated to locate the vortex phase singularity within a pixel.  Also, this method is less affected by the distortion of vortices orbiting near the edge of the trap.
\end{itemize}

\section{Reproducing the results}

The numerical computations in this work were carried out with version 1.5 of the Julia programming language, and the scripts that were used for generating the figures are included as separate supplemental material files. The figures may be reproduced by installing Julia, then unpacking {\bf gpvortex.tgz} to create the {\tt gpvortex} directory.  In that directory, run {\tt julia -e 'using Pkg; Pkg.instantiate()'} to download and install the necessary software libraries.  Then, for example, {\tt julia -l figone.jl} will run the code in the file {\tt figone.jl} that generates Fig.~1 of the main text.  This will load simulation results from a file {\tt orbit.jld2}, which can be reproduced by running {\tt orbit.jl} (this will take some time). Finally, the data can be inspected and other figures plotted using the command line. The scripts rely on the library {\tt Superfluids.jl} to do the numerical computations.  

\section{Supplemental Videos}

There are five included animations (animated gif file format). The Berry curvatures in the animations show the relative distribution only, and the absolute color scale changes slightly between frames. The axis labels are omitted from the movies for the sake of visual clarity. The first four animate the snapshot images shown in the main text. The animation SV5 shows the condensate ground state in a rotating frame as a function of the radial position of the vortex. The orbiting vortex transforms into an edge mode as the radius of its orbit approaches the Thomas--Fermi radius of the condensate. When the vortex is near the condensate edge, the vortex core localised kelvon quasiparticle mode hybridizes with the lowest energy surfon quasiparticle, resulting in the observed  ripples.
\begin{itemize}
    \item {\bf SV1.gif:} Imprinted vortex case corresponding to Fig.~1(a) and 1(d).
    \item {\bf SV2.gif:} GPE vortex case corresponding to Fig.~1(b) and 1(e).
    \item {\bf SV3.gif:} Moat vortex case corresponding to Fig.~2.
    \item {\bf SV4.gif:} One phase cycle of the displacement current in Fig.~3(a),(b),(c) and 3(d).
    \item {\bf SV5.gif:} Rotating ground state as function of the radial vortex position corresponding to Fig.~1(b) and (f).
\end{itemize}

%\bibliography{ref_list}
%\bibliographystyle{apsrev4-1}

\end{document}